\documentclass[conference]{IEEEtran}
\IEEEoverridecommandlockouts
\usepackage{textcmds}
\usepackage{url}
\usepackage{siunitx}
\usepackage{tikz}
\usepackage{booktabs}
\usepackage{multirow}
\usepackage{tabularx}
\usepackage{soul} 
\usepackage{cite}
\usepackage{graphicx}
\usepackage{amsmath,amssymb}
\usepackage{epstopdf}
\AppendGraphicsExtensions{.tif}
\usepackage{siunitx}
\ifCLASSINFOpdf
\else
\fi
\begin{document}
%
\title{Reproducible Physiological Features in Affective Computing: A Preliminary Analysis on Arousal Modeling\\

\thanks{\\
This work was supported by the Italian Ministry of Education and Research (MIUR) within the framework of the ForeLab Project (Departments of Excellence); by the PNRR—M4C2–Investment 1.3, Extended Partnership PE00000013 – ‘FAIR – Future Artificial Intelligence Research’, Spoke 1 ‘Human-centered AI’, funded by the European Commission under the NextGenerationEU programme; and by the ERC Starting Grant ScReeningData under grant number 101042407.
Andrea Gargano, Mimma Nardelli, and Enzo Pasquale Scilingo are with the Università di Pisa and Research Center \qq{E. Piaggio}.
Jasin Machkour and Michael Muma are with the Robust Data Science Group at Technische Universität Darmstadt.\\
Corresponding author: Andrea Gargano. Email: andrea.gargano@phd.unipi.it}
}

\author{\IEEEauthorblockN{1\textsuperscript{st} Andrea Gargano}
\IEEEauthorblockA{
\textit{Università di Pisa}\\
Pisa, Italy 
}

\and
\IEEEauthorblockN{2\textsuperscript{nd} Jasin Machkour}
\IEEEauthorblockA{
\textit{TU Darmstadt}\\
Darmstadt, Germany
}

\and
\IEEEauthorblockN{3\textsuperscript{rd} Mimma Nardelli}
\IEEEauthorblockA{
\textit{Università di Pisa}\\
Pisa, Italy 
}

\and
\IEEEauthorblockN{4\textsuperscript{th} Enzo Pasquale Scilingo}
\IEEEauthorblockA{
\textit{Università di Pisa}\\
Pisa, Italy 
}

\and
\IEEEauthorblockN{5\textsuperscript{th} Michael Muma}
\IEEEauthorblockA{
\textit{TU Darmstadt}\\
Darmstadt, Germany 
}

}

%


\maketitle
\begin{abstract}
In Affective Computing, a key challenge lies in reliably linking subjective emotional experiences with objective physiological markers.
This preliminary study addresses the issue of reproducibility by identifying physiological features from cardiovascular and electrodermal signals that are associated with continuous self-reports of arousal levels. 
Using the Continuously Annotated Signal of Emotion dataset, we analyzed 164 features extracted from cardiac and electrodermal signals of 30 participants exposed to short emotion-evoking videos. 
Feature selection was performed using the Terminating-Random Experiments (T-Rex) method, which performs variable selection systematically controlling a user-defined target False Discovery Rate. 
Remarkably, among all candidate features, only two electrodermal-derived features exhibited reproducible and statistically significant associations with arousal, achieving a 100\% confirmation rate. 
These results highlight the necessity of rigorous reproducibility assessments in physiological features selection, an aspect often overlooked in Affective Computing.
Our approach is particularly promising for applications in safety-critical environments requiring trustworthy and reliable white box models, such as mental disorder recognition and human-robot interaction systems.

\end{abstract}


%
\IEEEpeerreviewmaketitle


\section{Introduction}

In the Affective Computing field, detecting objective markers of human affect from physiological signals is an extensively studied problem \cite{Saganowski_2023}. 
Seminal works in the field, such as \cite{picard2001toward}, have demonstrated the potential of affect recognition approaches based solely on physiological signal analysis.
Since then, there has been sustained interest in developing physiology-based markers that enable the study of human emotional responses through objective physiological phenomena.
The analysis of physiological signals in emotion science is inherently motivated by the nervous system innervation of several body organs, particularly via its sympathetic and parasympathetic branches, both involved in the emotional response.
Thus, the goal is to indirectly characterize the relation between emotions and nervous system activity using minimally invasive and cost-effective physiological recordings and their analysis \cite{Saganowski_2023}.

Researchers have extensively used signal processing and feature extraction techniques to analyse various physiological signals, such as Autonomic Nervous System (ANS) correlates \cite{Kreibig_2010,Saganowski_2023}. 
Nevertheless, to the best of our knowledge, there has been limited investigation of the reproducibility of features extracted from physiological signals for the use in Affective Computing across different settings and applications \cite{Campbell2019, Giannakakis2022}. 
Indeed, a well-known issue in studies using classification techniques is the derivation of large physiology-based feature spaces that - after a refinement through feature selection stages - perform well on specific datasets but often fail to replicate under similar experimental conditions. 
However, for physiological patterns observed during emotionally salient events to be considered reliable and reproducible, they must demonstrate stability and consistency across studies \cite{Kreibig_2010}. 
This replication issue, defined as the \qq{replication crisis}, is a well-documented problem in scientific fields such as genomics and psychology \cite{baker20161}.

In this preliminary work, our aim is to advance and promote reproducible analysis of physiological markers in Affective Computing studies by utilizing a novel variable selection framework that rigorously controls the False Discovery Rate (FDR).
This work applies the Terminating-Random Experiments (T-Rex) selector \cite{machkour2025terminating, machkour2025da}, a new fast and open-source variable selection framework that controls a user-defined target FDR while maximizing the number of selected reproducible variables.
We employed it to analyze physiological features already used in the existing literature for the analysis of the self-reported intensity of experienced emotion (i.e., arousal), aiming to identify reproducible and statistically robust features that can enhance the reliability and effectiveness of Affective Computing applications. 
We used the Continuously Annotated Signal of Emotion (CASE) dataset \cite{sharmacase}, which provides synchronized recordings of both physiological signals and continuous self-assessed annotations of arousal and valence dimensions collected using an effective joystick-based interface \cite{sharma2017continuous}. We focused exclusively on the arousal annotations and a total of 164 features extracted from the electrocardiogram (ECG) and electrodermal activity (EDA) channels. 
We found that only 2 features consistently exhibited reproducibility and high statistical significance.
\subsubsection*{Organization}
Section~\ref{sec:methodology} introduces the CASE dataset, the physiological signal processing steps, and the computation of the feature space. It also describes the T-Rex variable selector for dependent features and the statistical analysis for assessing the reliability of the selected variables. 
The selected variables and the outcomes of the statistical models are presented in Section~\ref{sec:results}. The discussion and future directions are presented in Section~\ref{sec:conclusion}.

\section{Materials and Methods}\label{sec:methodology}

\subsection{The CASE Dataset}\label{sec:case}

The CASE dataset is an open source collection of physiological signals and continuous annotations gathered from a cohort of 30 healthy young adults (15 females, mean age $25.7\!\pm\! 3.1$ years; 15 males, mean age $28.6\! \pm\! 4.8$ years) while being exposed to 8 different emotion-arousing short video clip stimuli \cite{sharmacase, sharma2017continuous}.
The clips were purposely selected to evoke four distinct emotional states in the viewers (2 clips for each state): relaxation, fear, amusement, and boredom. 
The duration of each of the eight videos ranged from 119 to 197 seconds. 
Each participant saw the videos in a pseudo-randomized order to discard any potential bias induced by the presentation of the same sequence of stimuli. 
In addition, a two-minute-long blue screen video was played in-between two consecutive video stimuli as washout phase. 

We selected the CASE dataset because it provides moment-to-moment annotations of emotions. 
In fact, during each stimulation, participants continuously self-reported their affective state in terms of arousal and valence dimensions, according to Russell's Circumplex Model of Affect \cite{russell80}.
The emotional dimensions were rated on a continuous scale from $0.5$ to $9.5$ and collected using a custom emotion reporting interface, consisting of a joystick allowing to move and/or hold a pointer inside the bidimensional emotional space \cite{sharmacase}.
A graphical display of the valence-arousal plane, visually appearing on a section of the screen, provided a visual feedback of the continuous annotations to the participant. 
Experimenters specifically instructed participants to focus on the assessment of the momentary experienced emotion. To this aim, a training session of five short videos allowed users to familiarize themselves with the annotation task and the user interface prior to the data collection stage \cite{sharmacase}.
Statistically significant differences in the mean arousal and valence values for the four emotional conditions were reported in \cite{sharma2017continuous}, demonstrating the effective emotional induction in participants measured via the analysis of continuous annotations.

Concurrent recordings of several physiological measurements with continuous annotations occurred \cite{sharmacase}. 
The physiological signals and continuous annotations were acquired with sampling frequencies of \SI{1000}{\hertz} and \SI{20}{\hertz}, respectively. 
In this work, we analyzed the linearly interpolated arousal annotations of all eight emotion-eliciting videos, as well as the corresponding ECG and EDA signals. The linearly interpolated data, provided in \cite{sharmacase}, are a convenient data source in which latencies - possibly occurring due to lags in data acquisition and logging - were addressed through linear interpolation of the raw data. 
Specifically, we used as response the mean value of the self-assessed arousal level over the same temporal window of the physiological signals (see Sec.~\ref{sec:features}). 


\subsection{Physiological Signal Processing and Feature Extraction}\label{sec:features}

\begin{figure}[ht!]
    \centering
    \includegraphics[width=1\linewidth]{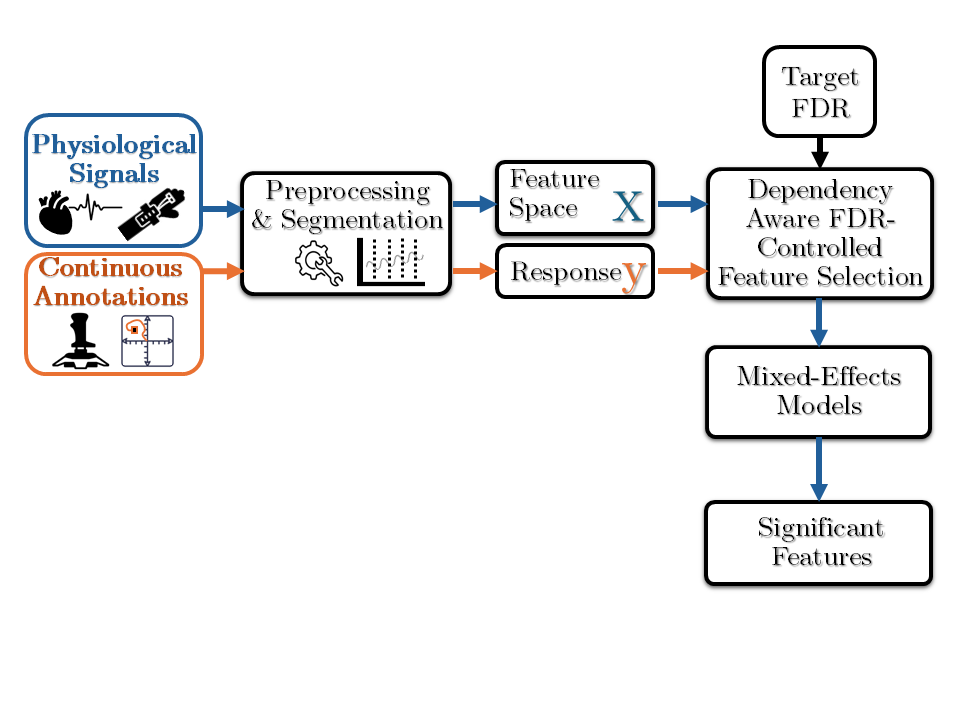}
    \vspace{-1.5cm}
    \caption{Block diagram of the proposed approach to discover reliable physiological features. The blue arrows show the path followed by the physiological features whereas the orange arrows is the path followed by the continuous annotations.}
    \label{fig:schema_da}
\end{figure}

As reported in Figure~\ref{fig:schema_da}, the first stage of our approach is the processing of the physiological signals, comprising the preprocessing and segmentation steps, followed by the feature extraction for the generation of the feature space.

As ECG preprocessing step, we analyzed the raw signals using Kubios HRV software (version 2.2) \cite{Tarvainen_2014} to detect R-peak locations. 
Subsequently, we derived the RR interval series, representing the time series of temporal intervals between successive R-peaks, the most investigated signal of cardiac dynamics in psycho-physiological studies.  
The EDA signal measures variations in skin conductance and consists of two different components - tonic and phasic - characterized by different time scales and physiological mechanisms. Therefore, for the EDA signal decomposition, we employed cvxEDA \cite{greco2015cvxeda}, a physiology-inspired and 
model-based convex optimization method, deriving the tonic component - Skin Conductance Level (SCL) - and the phasic component - Skin Conductance Response (SCR). 
The only pre-processing steps we applied to the raw EDA signals before using cvxEDA was their {Z-score} normalization, followed by a reduction of the sampling frequency by successive decimation from \SI{1000}{\hertz} up to \SI{50}{\hertz}, to aid model convergence \cite{greco2015cvxeda}. 
Additionally, we used the Sudo-Motor Neural Activity (SMNA) signal, an estimate of the ANS neural input underlying the SCR component, provided by cvxEDA.
Finally, all signals (i.e., RR, SCL, SCR, and SMNA) were segmented in windows of \SI{116}{\second}, based on the duration of the shortest signal recorded. Indeed, only the last \SI{116}{\second} of each signal per participant and video stimuli sessions were analyzed. 

From all signals, we extracted a total of 164 features: 132 from the RR signal, covering temporal, geometrical, frequency, nonlinear, and visibility-graph domains; 32 from the decomposition of the EDA signals into SCL, SCR, and SMNA, spanning temporal, frequency, and nonlinear domains; and 2 features computed by combining frequency domain features from RR and the sum of the SCL and SCR signals. A summary of the features extracted is reported in Table~\ref{tab:rr_eda_features}, where the last column displays the number of features for each combination of domain and signal type.

Time-domain analysis aims to capture both long- and short-term RR series variability in cardiovascular signals \cite{Task_force_1996, acharya2007advances} and standard temporal characteristics of the electrodermal components \cite{greco2015cvxeda}. For the HRV analysis, we extracted statistical indices from the RR-interval series to quantify global variability and short-term autonomic shifts. Similarly, geometrical-domain analysis of the RR intervals aims to quantify the variability of the signal based on characteristics of the empirical density \cite{Task_force_1996}. 
Regarding the EDA signal, we used central tendency and dispersion measures to quantify overall sympathetic-mediated responses (in the SCR and SMNA) and tonic baseline fluctuations (in the SCL). We computed these metric both over the entire observation window and as windowed averages, adapting the window to the specific transient nature of the electrodermal components \cite{Baldini_2022, greco2015cvxeda}.

Frequency-domain analysis decomposes each signal into oscillatory components associated with particular physiological rhythms. We first resampled the RR intervals at \SI{4}{\hertz} via piecewise cubic hermite interpolation, then applied Welch’s method (\SI{30}{\second} windows, 75\% overlap) to estimate HRV power in the low- and high-frequency bands, \SI{0.04}{\hertz}-\SI{0.15}{\hertz} and \SI{0.15}{\hertz}-\SI{0.4}{\hertz} respectively, decoupling sympathetic and parasympathetic influence on the cardiac dynamics. Analogously, we compute EDA spectral power (EDASymp) in the range \SI{0.045}{\hertz}-\SI{0.250}{\hertz} by summing tonic and phasic components, using both periodogram and Welch estimators \cite{posada-quintero_power_2016}. Moreover, we combined EDA and RR series frequency measures by computing the ratio between EDASymp and HF power, hence capturing the sympatho-vagal interplay by using reliable estimates of sympathetic and vagal dynamics \cite{ghiasi_assessing_2020}.

Nonlinear-domain analysis probes nonlinear interactions and dynamical complexity overlooked by other methods \cite{Sassi_2015}. 
For the HRV analysis, we reconstructed the phase-space trajectories to compute recurrence quantification analysis (RQA) metrics \cite{Webber_1994, riley2005tutorials, Marwan_2007} and entropy measures (SampEn \cite{richman2000}, FuzzyEn \cite{chen2007characterization}, DistEn \cite{li2015}). We also computed fractal and Detrended Fluctuation Analysis (DFA) indices, symbolic-dynamics features, and higher-order spectra characteristics. 
Similarly, we applied phase-space complexity measures (i.e., comEDA \cite{Nardelli_2022}, MComEDA \cite{Lavezzo_2024}) that quantify multiscale entropy for the characterization of EDA signal nonlinear dynamics.
We further analyzed RR intervals with the natural visibility graph \cite{Lacasa_2008}, 
extracting network‐based metrics for the characterization of the graph-transformation of the RR series.

We performed the feature extraction using MATLAB (R2023b The MathWorks, Inc.) software, employing custom and publicly-available algorithms. 
All 164 features were obtained from the respective physiological signals over the same temporal window as for the analyzed arousal annotations, ensuring temporal alignment between the predictors (i.e., the physiological features) and the response variable (i.e., the mean arousal).
The response variable 
was calculated as the mean value of the arousal signal over the last 116-second temporal window. 
As shown in Figure~\ref{fig:schema_da}, we adopted the standard statistical learning terminology: the mean arousal from the continuous annotations is the response variable, i.e., the $\mathbf{y}$ vector, with $N\!=\!240$ (8 videos per each of the 30 participants), and the physiological features form the predictor matrix $X \in \mathbb{R}^{N \times p}$, with $p\!=\!164$ in our work.


\begin{table*}[!t]
\centering
\caption{Features extracted from the RR series and the EDA signal components: SCL, SCR, SMNA}
\label{tab:rr_eda_features}
\scriptsize
\setlength{\tabcolsep}{4pt}
\renewcommand{\arraystretch}{1.3}
\begin{tabularx}{\textwidth}{@{}l l X |c}
\toprule
\textbf{DOMAIN} & \textbf{SIGNAL} & \textbf{FEATURES} & \textbf{NUMBER}\\
\midrule

\multirow{5}{*}{TEMPORAL} 
  & RR \cite{Task_force_1996,acharya2007advances} 
    & meanRR, stdRR, SDSD, RMSSD, NN50, pNN50, meanDER1, stdDER1, meanDER2, stdDER2, SkewRR, KurtRR & 12 \\
    \cline{2-4}
  & SCL \cite{Baldini_2022} 
    & mean, median, std, MAD; & 4\\ 
    &&Averaged over 20s-long windows: mean, median, std, MAD & 4 \\
    \cline{2-4}
  & SCR \cite{Baldini_2022} 
    & mean, median, std, MAD, N peaks, MaxPeak, AmpSum; & 7 \\
    &&Averaged over 5s-long windows: mean, median, std, MAD, AmpSum & 5\\
\cline{2-4}
  & SMNA \cite{Baldini_2022} 
    & mean, MaxPeak, N peaks, AmpSum & 4\\
\midrule
GEOMETRICAL 
  & RR \cite{acharya2007advances,Task_force_1996} 
    & TriRR. TINN 
    & 2\\
\midrule

FREQUENCY 
  & RR \cite{Task_force_1996} 
    & LF\_power, HF\_power, LF\_perc, HF\_perc, LF\_nu, HF\_nu, LF/HF, LF\_peak, HF\_peak & 9\\
    \cline{2-4}
  & SCL+SCR \cite{posada-quintero_power_2016, Baldini_2022} 
    & EDASymp, EDASymp\_db, EDASymp\_nu, EDASymp\_Welch, EDASymp\_db\_Welch, EDASymp\_nu\_Welch & 6 \\
    \cline{2-4}
  & SCL+SCR+RR \cite{ghiasi_assessing_2020} 
    & EDASymp/HF, EDASymp\_Welch/HF & 2\\
\midrule

\multirow{7}{*}{NONLINEAR} 
  &  
    & LPP \cite{Contreras_2006,Nardelli_2018}: \\
  & 
    & - for $M\in[1,10]$: SD1(M), SD2(M), SD12(M), $\rho$(M), P\_surf(M), SDRR(M); & 60\\
  & 
    & - given the $M$ values: AUC\_SD1, AUC\_SD2, AUC\_SD12, AUC\_rho\_RR, AUC\_P\_surf, AUC\_SDRR; & 6\\
    \cline{3-4}
  & RR
    & Fractal measures: FracDim \cite{Sevcik_2010}, HurstExp \cite{acharya2007advances}; & 2\\
    \cline{3-4}
  & 
    & DFA \cite{Peng_1995,acharya2007advances,Sassi_2015}: $\alpha1$, $\alpha2$; & 2\\
    \cline{3-4}
  & 
    & Symbolic dynamics \cite{Kurths_1995,Porta_2001,Sassi_2015}: v0, v2, c1v, c3v; & 4\\
    \cline{3-4}
  & 
    & Attention Entropy \cite{Yang_2023}: max-max, min-max, max-min, min-max, average. & 5\\
    \cline{2-4}
  & SCR 
    & ComEDA \cite{Nardelli_2022}, MComEDA \cite{Lavezzo_2024} & 2\\
\midrule

NONLINEAR: PHASE SPACE  
  & RR 
    & RQA \cite{Webber_1994,riley2005tutorials,Marwan_2007} : rec\_rate, det, avg\_diag, ratio, ent, lam, trap\_time, max\_len, mean\_rec\_time; & 9\\
    \cline{3-4}
  & 
    &SampEn \cite{richman2000}, FuzzyEn \cite{chen2007characterization}, DistEn \cite{li2015} & 3\\
\midrule

NONLINEAR: HIGHER‐ORDER SPECTRA 
  & RR \cite{acharya2007advances,schlogl2002time} 
    & Phase\_Entr, Mean\_Magn, Mean\_P, std\_P, N\_Bis\_Ent, N\_Bis\_Sq\_Ent, Sum\_log\_Amp, LL\_RR, LH\_RR, HH\_RR & 10\\
\midrule

GRAPH 
  & RR 
    & ShortPathLen, GlobClusterCoef \cite{Hou_2016}, mean LocalClusterCoef, mean Degree \cite{newman2018networks} & 4\\
\bottomrule
\end{tabularx}
\end{table*}

\subsection{The T-Rex Framework for Dependent Variables}\label{sec:trex_da}

Given the high degree of multicollinearity observed within the feature space, we adopted a Dependency-Aware variant of the T-Rex Selector (i.e., the T-Rex+DA+NN) that has shown effective FDR control in settings involving arbitrary overlapping
groups of highly correlated variables \cite{machkour2025nn}.
The T-Rex selector \cite{machkour2025terminating} is a fast FDR-controlling variable selection method for high-dimensional data, where the number of predictors $p$ may exceed the number of samples $N$. 
The T-Rex aim is to return an FDR-controlled set of selected variables, i.e., the features which are reliably related to the response. 
For doing so, the T-Rex framework comprises several steps, such as \cite{machkour2025terminating}:
\begin{itemize}
    \item Leveraging $K$ independent random experiments by employing L dummy predictors, sampled from the univariate standard normal distribution, and let them compete with the predictors in the feature space;
    \item Using forward variable selection algorithms 
    and early terminating them as soon as a given number $T$ (iteratively increased) of dummy variables is selected; 
    \item Returning variables whose relative occurrence across the $K$ random experiments exceeded a voting threshold $v$. The optimal values of $v$, $L$, and $T$ are jointly determined using an optimization algorithms that ensures the expected value of the false discovery proportion lays below the user-defined target threshold $\alpha$.
\end{itemize}

The extension of the T-Rex selector to its dependency-aware version (T-Rex+DA) was achieved by accounting for dependencies among variables, replacing the ordinary relative occurrences by dependency-aware relative occurrences \cite{machkour2025da}. 
In particular, a group design was devised to model the dependency structure within the predictors, assigning variables to groups depending on a correlation-based measure of similarity. 
A penalization term leverages the proposed group design and penalizes the relative occurrence of a given variable according to the resemblance with the relative occurrences of all the other variables lying within the same group. 
The parameter that controls the group sizes is automatically determined, being included in the automatic calibration algorithm that ensures FDR control.
In particular, the advantage of the T-Rex+DA+NN approach is that it incorporates a nearest neighbors (NN) penalization mechanism into the T-Rex+DA framework \cite{machkour2025nn} to manage overlapping groups of correlated variables. 

In this work, we leveraged the R package \textit{TRexSelector} \cite{trexpack}, using the T-Rex+DA+NN selector and setting the number of random experiments to 100 \cite{machkour2025terminating}. We experimentally evaluated values of $\alpha$ in the range [5\%, 25\%] with a step size of 5\%, and reported the significant features for $\alpha=10\%$.

\subsection{Statistical Analysis}

We evaluated the predictive utility of the selected features conducting statistical analysis based on linear mixed-effects models (LMER) (Fig.~\ref{fig:schema_da}).
Specifically, we fitted the linear relationship between the response and the selected features using a LMER model \cite{Bates_2015} and its robust version, the RLMER \cite{Koller_2016}. 
The choice of a robust model as comparison derives from the capability of robust approaches to better handle outliers and deviation from normality assumptions \cite{zoubir2018robust}.
The usage of mixed-effects model is motivated by the repeated-measures design of the analyzed data (see Sec.~\ref{sec:case}), using participant ID as a random intercept and the selected variables as fixed effects. 
For a general case where $s$ out of $p$ variables are selected, the model formula takes the form:
\begin{equation}
    \mathrm{Mean}_\mathrm{AR} \sim \mathbf{x_{1}} \ + \ ... \ +\  \mathbf{x_{k}} \ + \ ...\ + \ \mathbf{x_{s}} \ + \ (1 | \mathrm{ID})
\end{equation}
where $\mathbf{x_{k}}$ is the k-th selected variable and $k \in [1, s]$ spans the selected variables only. 

To account for multiple comparisons within each regression model, p-values were adjusted using the Benjamini-Hochberg procedure \cite{benjamini_controlling_1995}. 
For the computation of the p-values from the RLMER models, we used the degrees of freedom of the LMER model since they are not reported by the respective function \cite{Koller_2016}.
We performed the statistical analysis using R software (version 4.2).


\section{Results}\label{sec:results}

\begin{figure}[t!]
    \centering
    \includegraphics[width=0.5\linewidth]{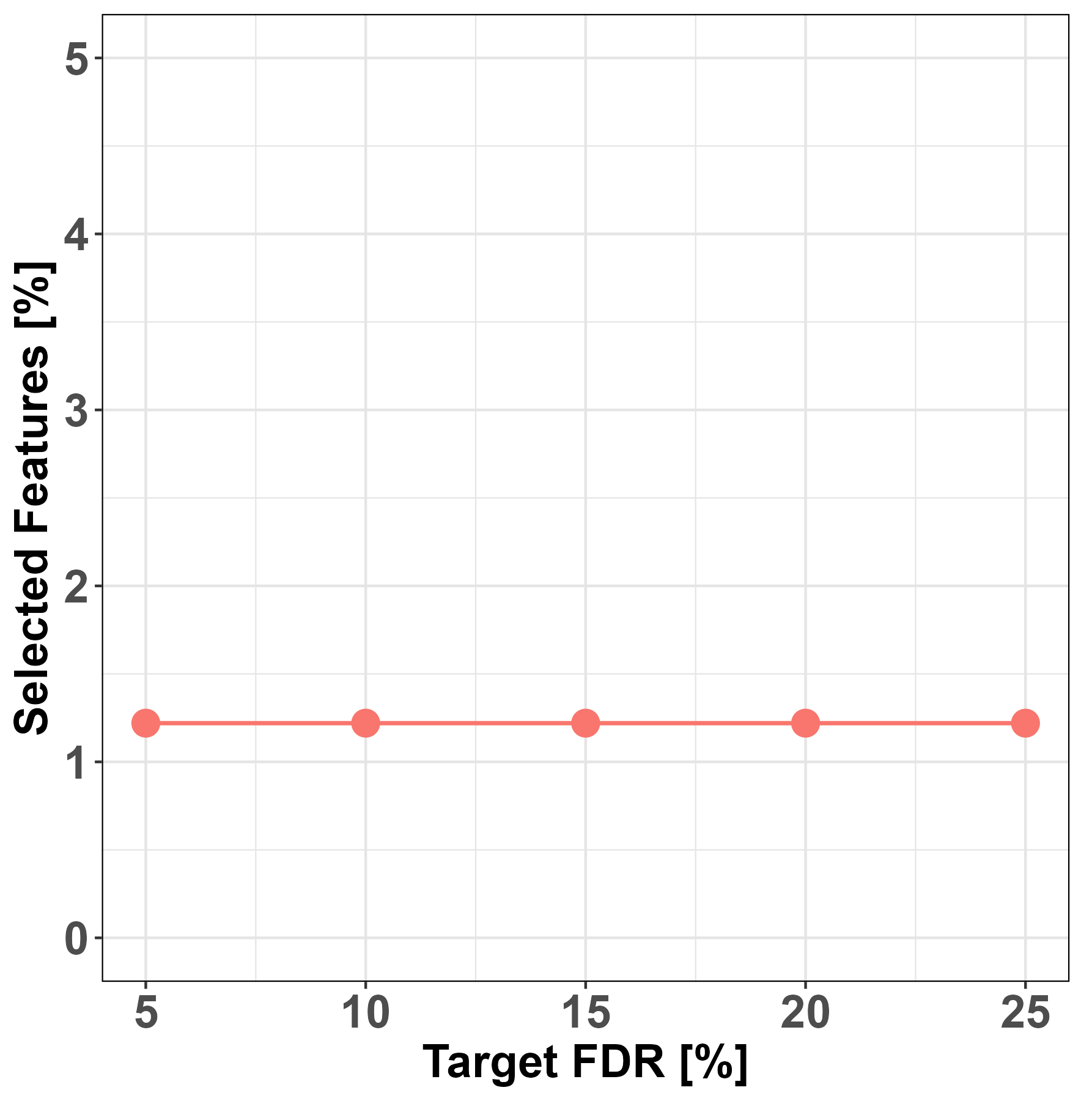}
    \caption{Percentage of selected features by varying values of the target FDR in the range [5\%, 25\%] with a step size of 5\%.}
    \label{fig:percentage}
\end{figure}

\begin{figure}[ht!]
    \centering
    \includegraphics[width=\linewidth]{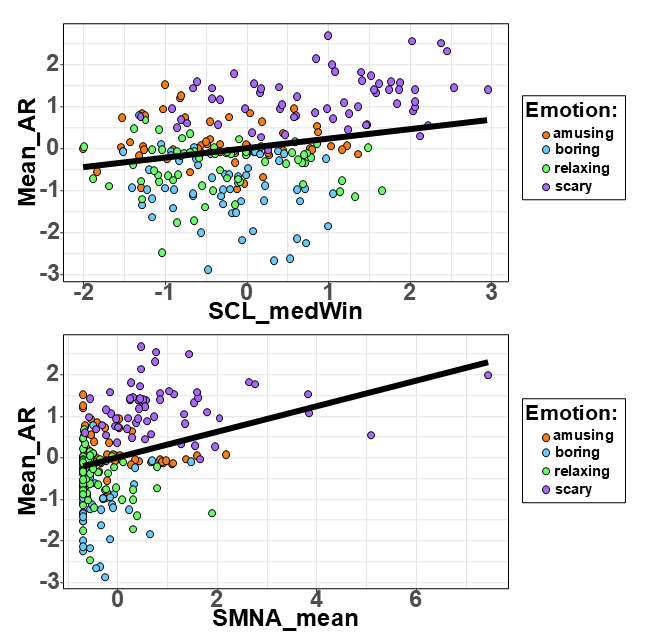}
    \caption{Representation of the estimated effects of the relation between the mean arousal level and (top) the averaged SCL median value within 20 s long windows, i.e. SCL\_medWin, and (bottom) the mean value of the SMNA, i.e., SMNA\_mean. 
    The slope of the regression line (in black) reflects the estimated effects of the LMER model. Different colours map the observed values to the four emotion classes, according to the legend on the right.}
    \label{fig:effects}
    \vspace{-0.5cm}
\end{figure}

Figure~\ref{fig:percentage} shows the percentage of selected features by setting different FDR target values as input parameters to the T-Rex+DA+NN selector. 
The percentage of selected features equals 1.21\% when the target FDR is set to 5\%, and remains the same even though the target values change in the range [5\%, 25\%]. Moreover, the selected features are stable, corresponding always to the same features.
In fact, among the 164 features tested, the T-Rex+DA+NN selector reliably identified only 2 features: the median of the averaged SCL signals calculated within non-overlapping 20-second segments ({SCL\_medWin}), and the mean value of the SMNA signal computed over the entire temporal window ({SMNA\_mean}). 
According to the outcome of both the LMER and RLMER models, these features were highly statistically significant, corresponding to a $100\%$ confirmation rate of significance among selected predictors.
More specifically, for the LMER model, the estimated fixed effect of {SMNA\_mean} is 0.309 ($\mathrm{SE}\!\!=\!\!0.065$) with an adjusted p-value$\ll\!\!0.001$ ($t\!=\!4.788$, p-value$=\!\!3.77e^{-6}$) whereas the estimated fixed effect of {SCL\_medWin} is 0.225 ($\mathrm{SE}\!\!=\!\!0.065$) with an adjusted p-value$<\!\!0.001$ ($t\!=\!\!3.481$, p-value$=\!\!5.96e^{-4}$). 
We observed similar results for the RLMER model. The estimated coefficient of {SMNA\_mean} equals 0.302 ($\mathrm{SE}\!\!=\!\!0.064$) with an adjusted p-value$\ll\!\!0.001$ ($t\!=\!\!4.709$, p-value$=\!\!5.30e^{-6}$) whereas the estimated fixed effect of {SCL\_medWin} is 0.225 ($\mathrm{SE}\!\!=\!\!0.065$) with an adjusted p-value$<\!\!0.001$ ($t\!=\!\!3.575$, p-value$=\!\!4.25e^{-4}$). 
In Figure~\ref{fig:effects}, the estimated effects for both features are represented as the slopes of the regression lines on top of the observed data. 
A detailed description of the summary statistics and performances for both tested models is provided in Table~\ref{tab:lmer_comparison}.

\begin{table}[ht!]
\centering
\setlength{\tabcolsep}{3pt}
\renewcommand{\arraystretch}{1.1}
\caption{Comparison of LMER and RLMER models for Mean Arousal Regression}
\label{tab:lmer_comparison}
\vspace{-0.3cm}
\begin{tabular}{@{}l c c @{}}
\toprule
 & \textbf{LMER} & \textbf{RLMER}\\
\midrule
\textbf{(Intercept)} 
  & -0.000 & -0.029 \\
  & (0.058) & (0.057)\\

\textbf{SMNA\_mean} 
  & 0.309*** & 0.302*** \\
  & (0.065) & (0.064)\\

\textbf{SCL\_medWin} 
  & 0.225*** & 0.229*** \\
  & (0.065) & (0.064) \\

\textbf{SD (Intercept:ID)} 
  & 0.027 & 0.000\\

\textbf{SD (Observations)} 
  & 0.894 & 0.866\\

\midrule
\textbf{N} & 240 & 240\\
\textbf{R\textsuperscript{2} Marg.} 
  & 0.206 & 0.215 \\
\textbf{R\textsuperscript{2} Cond.} 
  & 0.207 & - \\
\textbf{AIC} 
    & 645.7 & - \\
\textbf{BIC}
    & 663.0 & - \\
\textbf{ICC} 
    & 0.0 & - \\
\textbf{RMSE} 
    & 0.89 & 0.89 \\
\bottomrule
\end{tabular}
\end{table}

\section{Discussion and Conclusion}\label{sec:conclusion}
Our preliminary results underscore the critical importance of thoroughly evaluating reproducibility in Affective Computing, particularly in identifying robust physiological features that reliably map emotional dimensions. 
The selection of features derived from EDA is consistent with existing literature suggesting that EDA is particularly sensitive to the activity of the sympathetic nervous system and therefore closely associated with arousal states \cite{greco2015cvxeda, ghiasi_assessing_2020, Nardelli_2022, Baldini_2022}. 
Indexes of sympatho-vagal balance sensitive to arousing states (e.g., LF/HF) were probably not selected due to their sensitivity to the length of the temporal window.
However, our findings are inherently limited by the narrow set of the emotional states available, the small sample size, and the population tested. 
To further advance this line of research, we emphasize the need for larger-scale, high-quality studies involving substantially more participants and diverse populations. 
This would enable a comprehensive evaluation of broader sets of physiological features as well as additional channels under varied experimental conditions, increasing the likelihood of identifying a larger set of reproducible markers of emotional states \cite{Kreibig_2010}.
Testing this approach on broad publicly available datasets would not only enhance statistical power, but also improve generalization capability across contexts, demographics, and sensor modalities. 
Moreover, by applying and benchmarking reproducibility-aware feature selection frameworks, such as the FDR-controlling T-Rex selector, future work could establish a trustworthy and interpretable set of features with validated relevance to emotional processes. 
In this way, our preliminary work contributes to the development of reliable systems in the Affective Computing field. 
Reliable physiological markers of emotion will disclose new avenues for developing robust psycho-physiological white box models, particularly in high-trust, safety-critical areas such as mental disorder assessment and adaptive human-computer and human-robot interaction systems.

\bibliographystyle{IEEEtran}
\bibliography{IEEEabrv, biblio}

\end{document}